\documentclass[preprint]{aastex}

\usepackage{emulateapj5}
\usepackage{graphics}
\usepackage{psfig}

\def\93j{{SN~1993J}}
\def\R{{\sl ROSAT}}
\def\A{{\sl ASCA}}

\def\gs{\mathrel{\mathchoice {\vcenter{\offinterlineskip\halign{\hfil
$\displaystyle##$\hfil\cr>\cr\sim\cr}}}
{\vcenter{\offinterlineskip\halign{\hfil$\textstyle##$\hfil\cr
>\cr\sim\cr}}}
{\vcenter{\offinterlineskip\halign{\hfil$\scriptstyle##$\hfil\cr
>\cr\sim\cr}}}
{\vcenter{\offinterlineskip\halign{\hfil$\scriptscriptstyle##$\hfil\cr
>\cr\sim\cr}}}}}
\def\ls{\mathrel{\mathchoice {\vcenter{\offinterlineskip\halign{\hfil
$\displaystyle##$\hfil\cr<\cr\sim\cr}}}
{\vcenter{\offinterlineskip\halign{\hfil$\textstyle##$\hfil\cr
<\cr\sim\cr}}}
{\vcenter{\offinterlineskip\halign{\hfil$\scriptstyle##$\hfil\cr
<\cr\sim\cr}}}
{\vcenter{\offinterlineskip\halign{\hfil$\scriptscriptstyle##$\hfil\cr
<\cr\sim\cr}}}}}

\begin{document}

\title{X-Ray Detection of a Pre-Supernova Evolution for the SN~1993J Progenitor}

\author{Stefan Immler\altaffilmark{1},
Bernd Aschenbach\altaffilmark{2} \& Q. Daniel Wang\altaffilmark{1}}
\affil{$^1$Astronomy Department, University of Massachusetts, Amherst, MA 01003}
%\affil{immler@astro.umass.edu}
\affil{$^2$Max-Planck-Institut f\"ur extraterrestrische Physik,
Postfach 1312, 85741 Garching, Germany}

\shorttitle{X-Ray Detection of a Pre-Supernova Evolution for SN~1993J}
\shortauthors{Immler, Aschenbach \& Wang }

\begin{abstract}

We report on the first detection of a pre-supernova (SN)
evolution in the X-ray regime. The results are based on \R\ observations 
of \93j\ ranging from six days to five years after the outburst. 
The X-ray observations are used to probe the SN shell interaction with the 
ambient circumstellar matter (CSM). After exploring various scenarios that might
explain the observed X-ray lightcurve with a $t^{-0.27}$ rate of decline, 
we present a coherent picture in terms of the interaction 
of the SN shock front with the CSM deposited by the progenitor's stellar wind. 
During the observed period, the SN shell has reached a radius of $3\times10^{17}$~cm 
from the site of the explosion, corresponding to $\sim10^4$ years in the 
progenitors stellar wind history. Our analysis shows that the mass-loss rate of 
the progenitor has decreased constantly from $\dot{M} = 4\times10^{-4}$ 
to $4\times10^{-5}~{\rm M}_{\sun}~{\rm yr}^{-1} (v_{\rm w}/10~{\rm km~s}^{-1})$ 
during the late stage of the evolution.
Assuming a spherically symmetric expansion, the circumstellar matter density
profile is found to be significantly flatter ($\rho_{\rm csm} \propto r^{-1.63}$) 
than expected for a constant mass-loss rate and constant wind velocity 
profile ($r^{-2}$). The observed evolution either reflects a decrease in the 
mass-loss rate, an increase in the wind speed or a combination of both,
indicating that the progenitor likely was making a transition from the red 
to the blue supergiant phase during the late stage of its evolution.

\end{abstract}

\keywords{supernovae: individual (SN 1993J) --- stars: mass loss ---
X-rays: individual (SN 1993J) --- X-rays: ISM}

\section{Introduction}
\label{introduction}

The interaction of a supernova (SN) with circumstellar medium (CSM) 
produces a fast shock wave in the CSM and a reverse shock wave into the 
outer supernova ejecta.
Two characteristic regions of X-ray emission are produced by the interaction:
the forward shock wave in the CSM at $T \sim 10^4$ km s$^{-1}$ produces gas 
with $\sim 10^9$ K, and the reverse shock wave in the supernova ejecta
$\sim 10^3$ km s$^{-1}$ less than the forward shock produces gas with
$T\sim 10^7$ K. The reverse shock is formed where the freely
expanding supernova ejecta catches up with the CSM shocked by the blast wave. 
At early times, the reverse shock front is radiative and a dense, cool 
($T< 10^4$ K) shell can build up downstream from the radiating region 
(Chevalier \& Fransson 1994). The dense shell can absorb X-rays from the
reverse shock region and reprocess them to lower energies.

This scenario was supported by observations of Type IIb \93j\ with \R\
(Zimmermann et al.\ 1994, 1996) and with \A\ (Kohmura et al.\ 1994).
The early X-ray spectrum at day 6 was hard, with $T\sim 10^{8.5}$ K. At day $\sim 200$,
a softer component with $T\sim 10^7$ K dominated. The emergence of the softer 
spectrum could be attributed to the decreased absorption by a cool shell
(Fransson, Lundqvist \& Chevalier 1996; hereafter: FLC96). 
%Recent observations of SN~1999em 
%with \C\ starting four days after the outburst also show that the X-ray 
%spectrum becomes increasingly softer with time (Pooley et al.\ 2001). 
%No other X-ray supernova has been well observed at such early 
%times\footnote{a complete list of X-ray SNe and references can be found at
%http://xray.astro.umass.edu/sne.html}. 

\begin{figure*}[t!]
\centerline{ {\hfil\hfil
\psfig{figure=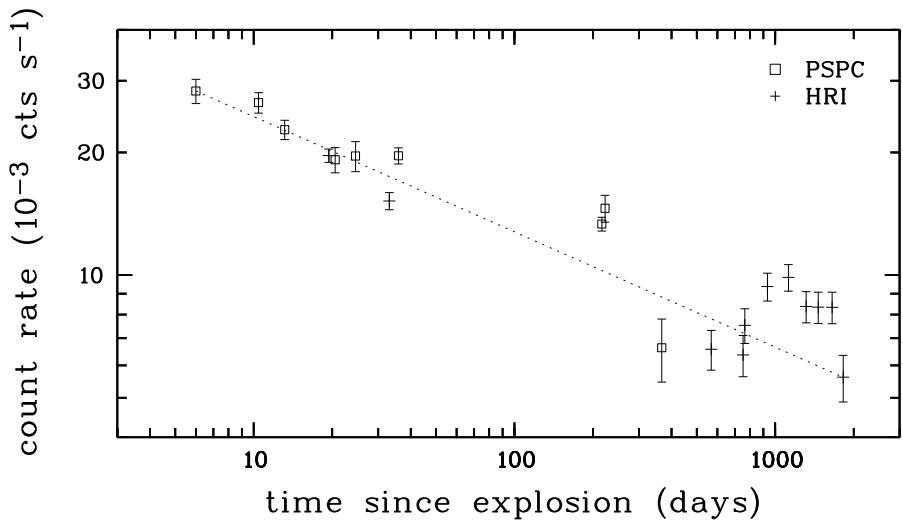,width=12cm,clip=}
\hfil\hfil} }
\caption{\R\ (0.5--2~keV band) X-ray lightcurve of \93j. Boxes mark PSPC data, 
HRI data are indicated by crosses. Error bars are $1\sigma$ statistical errors.
The dotted line illustrates a $t^{-0.27}$ rate of decline.
\label{f1}}
\end{figure*}

\section{Theoretical Background}
\label{theory}

The thermal X-ray luminosity $L_{\rm x}$ of the shock heated plasma 
is expressed by the product of the emission measure, EM, and the cooling 
function, $\Lambda(T, Z, \Delta E)$, where $T$ is the plasma temperature,
$Z$ represents the elemental abundance distribution, and $\Delta E$ is 
the X-ray energy bandwidth. For spherically symmetric conditions 
${\rm EM} = 4\pi \int_{R_0}^{R} n_{\rm e}^2 r^2 dr$, where $r$, the radial coordinate, 
runs from $R_0$ to $R$, the current outer boundary of the shocked matter which
has an electron density $n_{\rm e}$. Assuming a constant supernova shock wave speed 
$v_{\rm s}$, $R=v_{\rm s}t$, where $t$ represents the time elapsed since the 
explosion. If the ambient matter density $\rho$ is dominated by a wind blown 
by the progenitor star of the supernova the continuity equation requires 
${\dot{M}}=4\pi r^2 \rho v_{\rm w}$, with ${\dot{M}}$ the mass loss rate 
and $v_{\rm w}$ the wind speed. The X-ray luminosity can then 
be expressed as $L_{\rm x} = \frac{1}{4\pi} \Lambda(T) 
\times ({\dot{M}}/{v_{\rm w}})^2 \times (v_{\rm s}~t)^{-1}$.
We can hence use the X-ray luminosity to measure the ratio 
${\dot{M}}$/$v_{\rm w}$ of the progenitor star.

Each X-ray measurement at $t$ is related to the corresponding 
distance from the site of the explosion.
This site had been reached by the wind at a time depending on $v_{\rm w}$,
or the age of the wind $t_{\rm w} = t v_{\rm s}/v_{\rm w}$. 
Usually, $v_{\rm s} \gg v_{\rm w}$ so that with $t$ only a few 
years we can look back in time quite an appreciably large time span 
in the evolution of the progenitor's wind.
Assuming that $v_{\rm w}$ did not change over $t_{\rm w}$ we can even 
directly measure the mass loss rate back in time. Whether 
$v_{\rm w} = {\rm const}$ is a reasonable assumption will be discussed below.
Integration of the mass-loss rate along the path of the expanding shell gives 
the mean density inside a sphere of radius $r$. For a constant wind velocity 
$v_{\rm w}$ and mass-loss rate $\dot{M}$, a 
$\rho_{\rm csm} = \rho_0 (r/r_0)^{-s}$ profile with $s=2$ is expected.

After the expanding shell has become optically thin, it is expected that 
emission from the SN ejecta itself, heated by the reverse shock, dominates 
the X-ray output of the interaction regions due to its higher emission
measure and higher density. For a uniformly expanding ejecta
the density structure is a function of its expansion velocity,
$v$, and the time after the explosion, $t$: 
$\rho_{\rm sn}  = \rho_0 (t/t_0)^{-3} (v/v_0)^{-n}$ with $\rho_0$ 
the ejecta density at time $t_0$ and velocity $v_0$ (FLC96). 
For a red supergiant progenitor, the power-law is
rather steep with index $n\sim20$ (Shigeyama et al.\ 1994; Baron, Hauschildt 
\& Branch 1994; Suzuki \& Nomoto 1995; FLC96).
For constant $n$, the radius of the discontinuity surface between the 
forward and the reverse shock evolves in time $t$ with 
$R_{\rm c} \propto t^m$ where $m=(n-3)/(n-s)$ is the deceleration parameter.

\section{X-Ray Observations and Analysis}
\label{obs}

\93j\ in M81 was observed in 5 individual pointings with the
Position Sensitive Proportional Counter (PSPC) and in 10 individual pointings
with the High Resolution Imager (HRI) onboard \R\ (Tr\"umper 1983).
Total integration times with the PSPC and HRI instruments are 52.2~ks and 140.3~ks,
respectively. Assuming an initial explosion on March 28, 1993 (Ripero 1993),
the observations cover a period of 6 to 1\,821 days after the outburst.
A detailed description of the data calibration and analysis can be found in 
Immler \& Wang \cite{Immler01}.

In order to investigate the rate of decline for \93j, we constructed a 
combined PSPC+HRI lightcurve, which is presented in Fig.~\ref{f1}. 
HRI data were binned into 11 continuous observation blocks with 
5--20~ks exposure time each to obtain satisfactory counting statistics.
PSPC data were binned into 9 intervals with 2--18~ks integration time each. 
The time-dependent background was determined by normalizing the total 
background map of the HRI and PSPC images according to 
the total source-removed count rate in each exposure interval.
Background subtracted source counts were extracted within the 90\% radii
around the fixed position of \93j. In order to reduce background due to 
UV emission and cosmic rays, only HRI PI channels 2--10 were used.

An effective 0.5--2~keV band cooling function of 
$\Lambda=3\times10^{-23}$ ergs~cm$^{-3}$ s$^{-1}$ is adopted for an 
assumed optically thin thermal plasma with a temperature of $10^7$~K
(Raymond, Cox \& Smith 1976).
The equivalent count rate to (unabsorbed) flux conversion factor 
is $4 \times 10^{-11} $$({\rm~ergs~cm^{-2}~s^{-1})/(counts~s^{-1}})$
for a Galacti column density of 
$N_{\rm H}=4\times10^{20}~{\rm cm}^{-2}$ (Dickey \& Lockman 1990).
\R\ HRI and PSPC count rates were converted using a factor of 3 appropriate
for the assumed source spectrum (cf. Immler \& Wang 2001).
HRI count rate to flux conversion factors as a function of spectral
model parameters are given in Fig.~5 in Wang, Immler \& Pietsch (1999). 
The uncertainty of the HRI (PSPC) conversion factors for optically thin
thermal spectra with temperature in the range $10^{7}$--$10^{8.5}$~K 
(cf. \S\ref{introduction}) is 22\% (19\%). An increase in absorbing column 
from $10^{20}$ to $10^{21}~{\rm cm}^{-2}$ increases the unabsorbed source fluxes by
$\sim15\%$. Assuming a 0.86~keV thermal Bremsstrahlung spectrum (corresponding
to a temperature of $10^{7}$~K) instead of a Raymond \& Smith thermal plasma
reduces the HRI (PSPC) conversion factor by 4\% (1\%).

\section{Discussion}
\label{discussion}

\subsection{X-ray emission from the shock-heated stellar wind}
\label{wind}
We will first discuss the observed X-ray lightcurve in the
context of the shock-heated stellar wind model and explore other scenarios
later. In order to derive the stellar wind age prior to the outburst,
an initial wind velocity of $v_{\rm w}=10$~km~s$^{-1}$ and shock front 
velocity of $v_{\rm s}=19\,900$~km~s$^{-1}$ are assumed (Bartel et al.\ 1994).
The choice for the stellar wind speed is justified by the fact that \93j\
is a Type IIb SN with a massive ($\sim15M_{\odot}$ ZAMS) red supergiant progenitor
(e.g. Podsiadlowski et al.\ 1993). Typical outflow velocities for these progenitors 
are in the range of $3 \ls v_{\rm w} \ls 30$~km~s$^{-1}$, peaking at 
$\sim10$--$15$~km~s$^{-1}$ (e.g. Barnbaum, Kastner \& Zuckermann 1991).

Since the shock front catches up with the wind deposited by the 
progenitor with a speed of up to $\sim2\,000\times$
larger, the interaction front can be used to probe the stellar wind history
over a period of $\sim10^4$ years. Our X-ray observations can hence be
used as a `time machine' to directly measure the CSM density during the last
stages of the stellar evolution. 

\begin{figure*}[t!]
\psfig{figure=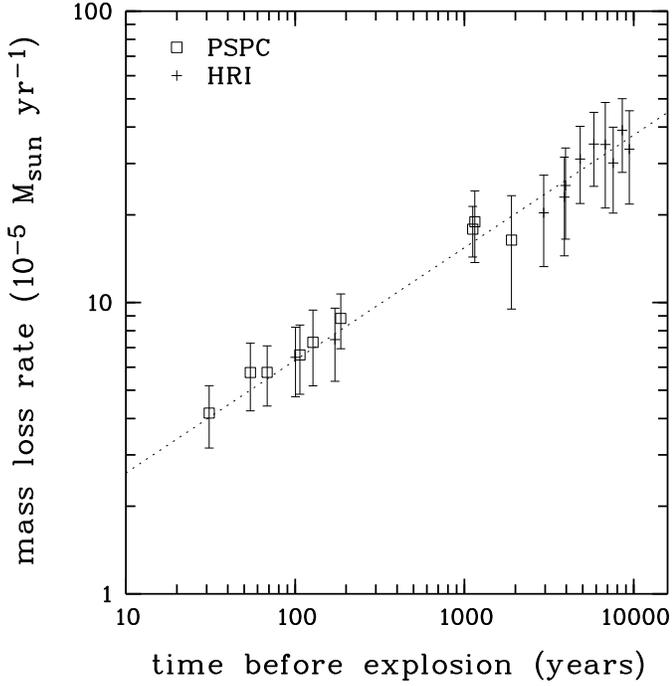,width=9cm,clip=}
\caption{Mass-loss rate history of the \93j\ progenitor.
PSPC data are marked by boxes, HRI data are indicated by crosses.
\label{f2}}
\end{figure*}

\begin{figure*}[h!]
\psfig{figure=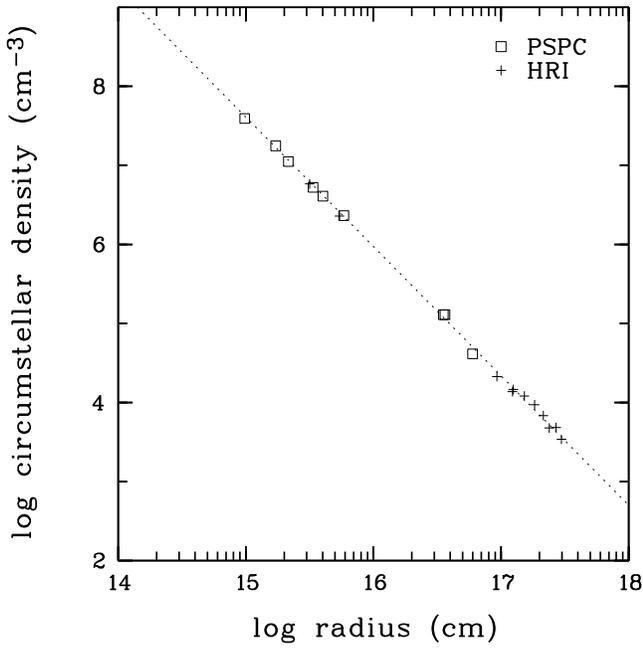,width=9cm,clip=}
\caption{Circumstellar density profile as a function of SN shell expansion radius.
The solid line gives the best-fit CSM profile of $\rho_{\rm csm} \propto r^{-1.63}$
to the PSPC (boxes) and HRI (crosses) data points.
\label{f3}}
\end{figure*}

Fig.~\ref{f2} illustrates the change in mass-loss rate as a function of the 
stellar wind age. It can be seen that the mass-loss rate constantly 
declined from $\dot{M} = 4\times10^{-4}$ to 
$4\times10^{-5}~{\rm M}_{\sun}~{\rm yr}^{-1} (v_{\rm w}/10~{\rm km~s}^{-1})$ 
just prior to the explosion. Integration of the mass lost by the progenitor 
along the shock front gives the CSM density at the given interaction radius 
from the SN site. The CSM density distribution is illustrated in Fig.~\ref{f3}. 
It is found that the CSM density profile is best described
by a power law $\rho_{\rm csm} \propto r^{-s}$ with index 
$s=1.63\pm0.02$, assuming a spherically symmetric expansion.
This is significantly flatter than expected for a constant mass-loss rate 
and constant wind velocity profile ($r^{-2}$). The results confirm previous
evidence for a rather flat CSM density profile during the first year 
($s\sim1.5$, van Dyk et al.\ 1994; $1.5 \ls s \ls 1.7$, FLC96; 
$s=1.66^{+0.12}_{-0.25}$, Marcaide et al.\ 1997).

The observed change in the pre-SN history could be due to variations in one
of the following parameters:

$\bullet$~Variations in velocity and/or temperature of the shocked CSM may have 
caused changes in the X-ray output of the shocked region. This model was proposed
to explain the radio lightcurve of \93j\ (Fransson \& Bj\"ornsson 1998).
In fact, evidence for a $\ls20\%$ decrease in the expansion velocity over the 
first $\sim4$ years was reported (Marcaide et al.\ 1997; Bartel et al.\ 2000).
This change in shock expansion velocity, however, would only lead to a 
$<10\%$ increase in the mass-loss rate and cannot account for the
observed order-of-magnitude change. Also, as the \R\ soft X-ray band cooling 
function for optically thin thermal plasma emission is not very sensitive to the 
plasma temperature ($\Lambda(T) \propto T^{0.5}$; e.g. Raymond, Cox \& 
Smith 1976), a change in temperature from $10^{7}$~K to a few $\times 10^{11}$~K
would be required to explain the observed change in the mass-loss rate history. 
Such drastic temperature variations have neither been observed for SNe nor 
have they been put forward by models describing the CSM interaction. 
Therefore, variations in either the shock expansion velocity or in the 
temperature accounting for the inferred evolution can be ruled out.

$\bullet$~A different scenario would be a non-spherically symmetric geometry 
caused by a binary evolution of the progenitor. Observational evidence 
that the progenitor of \93j\ was a stripped supergiant in a binary system has
been presented based on the visual lightcurve, which cannot be described
by a single, massive star alone (e.g. Podsiadlowski et al.\ 1993).
In addition, ``double-horned'' emission line profiles indicate the presence
of a flattened or disk-like expanding shell (Matheson et al.\ 2000a; 2000b).
The X-ray lightcurve also shows some significant deviations from the long-term
$t^{-0.27}$ behavior (cf. Fig.~\ref{f1}). Such deviations have also been
observed for SN 1979C in the radio regime (Montes et al.\ 2000) and were 
discussed in the context of a self-colliding binary stellar wind model leading 
to the formation of a multiple shell-like CSM profile (Schwarz \& Pringle 1996). 
X-ray observations of SN~1979C, however, were not suited to address these
questions (Immler, Pietsch \& Aschenbach 1998; Kaaret 2001). 
Similarly, a binary model for \93j\ cannot be challenged by the \R\ data alone 
and no information about the long-term radio lightcurve of \93j\ is available.

$\bullet$~Alternatively, the observed evolution is caused by changes in the 
wind parameters. The X-ray lightcurve implies $\rho_{\rm csm} \propto r^{-s}$ 
with $s=1.63$, or, using the continuity equation, 
$\dot{M}$/$v_{\rm w}  \propto r^{2-s} \propto r^{0.37}
\propto t_{\rm w}^{0.37}$. The ratio $\dot{M}$/$v_{\rm w}$ therefore decreases 
with time approaching the date of the explosion, indicating either a decrease 
in the mass-loss rate, an increase in the wind speed or a combination of both. 
It is interesting to compare typical wind parameters for red and blue supergiants, 
which are on the same evolutionary track for massive ($\gs15M_{\odot}$ ZAMS) 
stars (e.g. Salasnich, Bressan \& Chioso 1999).
Whereas the mass-loss rate at $t_{\rm w}\sim10^4$ years 
($\dot{M} = 4\times10^{-4}~{\rm M}_{\sun}~{\rm yr}^{-1} (v_{\rm w}/10~{\rm km~s}^{-1})$)
is typical for a red supergiant 
(e.g. $\dot{M} = 3\times10^{-4}~{\rm M}_{\sun}~{\rm yr}^{-1}$ for HD 179821, 
Jura, Velusamy \& Werner 2001; Chin \& Stothers 1990;
Schaller et al.\ 1992), the observed change in ratio $\dot{M}$/$v_{\rm w}$ 
by an order of magnitude at $t_{\rm w}\sim30$ years could indicate a transition 
between the red and blue supergiant phase due to an increase in effective 
temperature of the star. Blue supergiants are known to have 
significantly lower mass-loss rates and higher wind velocities
($\dot{M}\sim10^{-6}~{\rm M}_{\sun}~{\rm yr}^{-1}$,
$v_{\rm w}\sim500$--$1000~{\rm km~s}^{-1}$; e.g. Kudritzki et al.\ 1999).
This scenario for the evolution of the \93j\ progenitor would have some
interesting similarities with that of SN~1987A, whose progenitor 
completely entered the blue supergiant phase after significant mass-transfer 
to a companion (Podsiadlowski et al.\ 1993).

Whereas our X-ray data alone cannot give information on whether $\dot{M}$ or 
$v_{\rm w}$ have effectively changed during the late stage of the progenitor's
evolution, there is a clear difference between these possibilities
regarding the kinetical energy of the stellar wind:
using $\rho_{\rm csm} \propto r^{-1.63}$, the kinetical wind energy decreased
by a factor of 13 for a changing $\dot{M}$ during the observed 
$\Delta t_{\rm w} \sim 10^4$ years. In the scenario of a changing $v_{\rm w}$ 
only, an increase in kinetical wind energy by a factor of $10^{2.2}$ is expected. 
Future stellar evolutionary models might give answers as to which of the
two possibilities is more likely.

\subsection{X-ray emission from the shocked-heated SN ejecta}
\label{ejecta}
Let us explore the scenario that the X-ray emission is dominated by the ejecta,
heated by the reverse shock. A necessary condition for the ejecta accounting for
the observed X-ray emission is that the absorption by the post-shock gas is low.
In this model, an ejecta density structure of
$\rho_{\rm sn} = \rho_0 (t/t_0)^{-3} (v/v_0)^{-n}$ is expected
(FLC96). The total X-ray output of both the
reverse-shock heated ejecta and the shocked stellar wind in the forward shock
is then given by $L_{\rm x} \propto (\dot{M}/v_{\rm w})^2 \times (t/t_0)^{3-2s}$ 
in the case where $n$ is large. In fact, using our inferred X-ray rate of 
decline of $L_{\rm x} \propto t^{-0.27}$ is entirely consistent with the 
above $L_{\rm x} \propto t^{3-2s} \propto t^{-0.24}$ for $s=1.63$.
The assumption that $n$ must be large is hence confirmed by our data
and has also been concluded from radio and optical observations of \93j\
($n\sim20$--$30$; Shigeyama et al.\ 1994; Baron, Hauschildt \& Branch 1994; 
Suzuki \& Nomoto 1995; FLC96).
It is important to note that if the density gradient for the ejecta is large,
X-rays from the reverse shock must be heavily absorbed. Whereas variations
in temperature of the ejecta can be ruled out to explain the observed X-ray 
lightcurve (cf. \S\ref{wind}),
a change in absorption from initially $10^{22}~{\rm cm}^{-2}$ to the
Galactic column of $4\times10^{20}~{\rm cm}^{-2}$ could account for the
observed decrease in source flux (cf. Fig.~5 in Wang, Immler \& Pietsch 1999). 
The signature of high absorption, however, is absent in all \R\ PSPC spectra of \93j,
which are consistent with the Galactic foreground absorption
($N_{\rm H}=5.3\pm1.7\times10^{20}~{\rm cm}^{-2}$, Zimmermann et al.\ 1994, 
1996). The most likely scenario is hence that the emission from the
reverse shock is completely absorbed and that the observed soft X-rays 
are only due to the shocked CSM of the progenitor wind.

Based on the modeling of the early X-ray lightcurve of \93j, Fransson, Lundqvist \& 
Chevalier (FLC96) concluded that the X-ray emission during the first months 
originates from the interaction of the SN shock with the circumstellar medium.  
As the reverse shock begins to penetrate the cool shell it is expected to
contribute an increasing fraction to the total X-ray output. This predicted rise 
of the \R\ band X-ray flux after $\sim100$ days (as illustrated in Figs.~8 and 
10 in FLC96) due to the emerging ejecta is not observed with \R. Instead, the 
overall long-term X-ray lightcurve (Fig.~\ref{f1}) is declining with a power-law 
close to the initial measurements, which were consistently explained as a result 
of the interaction of the SN shell with the ambient CSM.

\section{Conclusions}
\label{conclusions}
We present the first detection of a pre-SN evolution in X-ray,
based on long-term monitoring of \93j\ with \R\ over a period of 5 years.
The data are fully consistent with a description in the context of the 
SN shock interacting with the CSM blown off by the progenitor's stellar wind.
From the X-ray rate of decline $L_{\rm x} \propto t^{-0.27}$ we infer a CSM 
profile $\rho_{\rm csm} \propto r^{-1.63}$, which is significantly flatter 
than expected for a constant mass-loss rate and constant wind velocity 
profile ($r^{-2}$). The observations cover an epoch of $\sim 10^4$ 
years in the progenitor's stellar wind history.
During this period, the mass-loss rate of the progenitor has decreased 
constantly from $\dot{M} = 4\times10^{-4}$ 
to $4\times10^{-5}~{\rm M}_{\sun}~{\rm yr}^{-1} (v_{\rm w}/10~{\rm km~s}^{-1})$ 
just prior to the explosion.
The most likely explanation for this pre-SN evolution is either an 
increase in wind speed, a decrease in mass-loss rate or a combination of both,
indicative that the progenitor star was undergoing a transition from the red 
to the blue supergiant phase. The data demonstrate the scientific potential of 
long-term X-ray monitoring of SNe as an important diagnostical tool to probe 
the CSM interaction and the evolution of SN progenitors. 

\acknowledgments

This research made use of various online services and databases,
e.g. ADS, HEASARC, NED, and the \R\ data archive at MPE. 
The project is supported by NASA under the grants NAG 5-8999 and NAG5-9429.

%%%%%%%%%%%%%%%%%%%%%%%%%%%%%%%%%%%%%%%
%% if Figures are included:
%%%%%%%%%%%%%%%%%%%%%%%%%%%%%%%%%%%%%%%

\end{document}